\documentstyle[twocolumn,prl,aps,graphicx]{revtex} % twocolumn format
\newcommand{\ksla}{ \parbox[b]{0.6em}{$k$} \hspace{-0.55em}
                         \parbox[b]{0.55em}{ \raisebox{-0.2ex}{$/$}}}
\newcommand{\be}{\begin{eqnarray}}
\newcommand{\ee}{\end{eqnarray}}

\begin{document}
\draft
\title{QCD Signatures of Narrow Graviton Resonances
in Hadron Colliders}

\author{J.~Bijnens$^a$, P.~Eerola$^b$, M.~Maul$^a$, A.~M{\aa}nsson$^a$, T.~Sj\"ostrand$^a$}
\address{$^a$ Department of Theoretical Physics, Lund University,
         S\"olvegatan 14A, S - 223 62 Lund, Sweden}
\address{$^b$ Department of Elementary Particle Physics, Lund University,
 Professorsgatan 1, S - 223 63 Lund, Sweden}
\date{\today}
\maketitle

\begin{abstract}
We show that the characteristic $p_\perp$ spectrum yields
valuable information for the test of models for the 
production of narrow graviton resonances in the TeV range 
at LHC. Furthermore, it is demonstrated that in those scenarios
the parton showering formalism agrees with the
prediction of NLO matrix element calculations.  

\end{abstract}
\narrowtext

\section{Introduction}
The search for particles beyond the Standard Model is one
of the key issues of the upcoming ATLAS and CMS experiments at LHC. If these
particles are generated from quarks and gluons, characteristic
signatures are expected from genuine QCD effects such as parton
showering. As a specific example we study the narrow graviton resonances
predicted by the Randall-Sundrum model \cite{Randall:1999ee}. As 
opposed to the concept of Large Extra Dimensions 
\cite{Arkani-Hamed:1998rs}, where a continuous spectrum
of Kaluza-Klein states is predicted, the Randall-Sundrum model
predicts a series of narrow heavy graviton resonances. 
Both models lead to a modification of the gravitation potential
at small distances $R$. While in case of two large extra dimensions
the precision reachable at LHC ($R=4.5 \;{\rm\mu}$m \cite{Antoniadis:2000vd})
is roughly comparable to present mechanical experiments
($R=218\; {\rm \mu}$m \cite{Hoyle:2000cv}), signatures for the Randall-Sundrum
type of gravitons can only be seen in TeV-scale collider experiments.
\newline
\newline
Recently, in the context of the Randall-Sundrum model the signatures of  
narrow graviton resonances at TeV-scale have been studied in 
Ref.~\cite{Allanach:2000nr}. Here the processes $gg \to G \to e^+e^-$ and
$q\bar q \to G \to e^+e^-$ are of interest as their
leptonic final states provide a clean and simple way to identify
the heavy graviton resonance $G$. The main experimental signature for such
a spin-2 graviton resonance is the characteristic angular distribution
of the produced $e^+e^-$ pair. As the extraction
of the angular distribution is quite difficult
even with high luminosity, further characteristic signatures
are desirable to make conclusive statements.
\newline
\newline
One possibility of  such a complementary signature
is the $p_\perp$ spectrum 
($p_\perp$ is the transverse momentum with respect to the 
beam direction) 
of the reconstructed
graviton resonance $G$. The production mechanism is given by a characteristic
mixture of $gg$ and $q\bar q$ fusion processes. 
As those processes are highly energetic,
the initial-state partons will radiate off a large amount of other partons,
leading to a $p_\perp$ spectrum which is different for
$gg$ and $q\bar q$ initial states. Especially, the larger color charge of the 
gluon will lead to more radiation and a
larger average $p_\perp$ in the former process.
So by studying the $p_\perp$ spectrum,
one can check the ratio of $q\bar q$ and $gg$ events producing the resonance
in question and compare that to the prediction of a certain model,
in our case the Randall-Sundrum model.   

\section{Physical Subprocesses}
\begin{figure}[tb]
\begin{center}
\includegraphics[width=8cm]{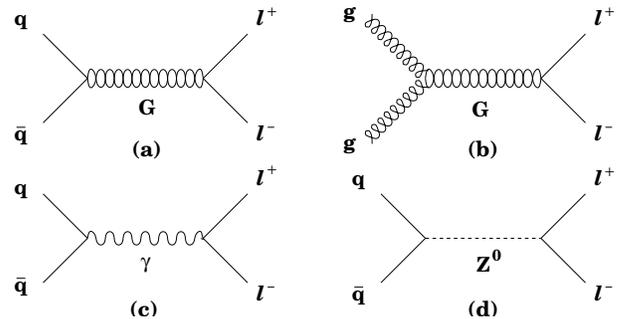}
\end{center}
\caption{Signal and Standard Model 
background processes for the graviton resonance
production.}
\label{graphs} 
\end{figure}

The partonic subprocesses of interest to discover the narrow graviton
resonance $G$ are $gg \to G \to l^+l^-$ and
$q\bar q \to G \to l^+l^-$, as depicted in Fig.~\ref{graphs} a,b.
The relevant Standard Model  background is the   $l^+l^-$ pair
production from virtual $Z^0$ bosons and photons, see Fig.~\ref{graphs} c,d, 
and their interference. The interference with the narrow graviton resonance
can be completely neglected since the mass of the graviton, if existent, has
to lie far above the $Z^0$ mass. 
%
%
%%%%%%%%%%%%%%%%%%%%%%%%%%% Figure 1 %%%%%%%%%%%%%%%%%%%%%%%%%%%%%%%%%%%%%%
%%%%%%%%%%%%%%%%%%%%%%%%%%%%%%%%%%%%%%%%%%%%%%%%%%%%%%%%%%%%%%%%%%%%%%%%%%%
%
%
For the cross sections (a) and (b) of the processes depicted 
in Fig.~\ref{graphs}
we reproduce in accordance with Ref.~\cite{Giudice:1999ck,Han:1999sg}:
\begin{eqnarray}
\frac{d\sigma^{(a)}}{dt} &=& \frac{1}{N_c}\frac{\kappa^4s^2}{2048\pi}
\frac{1+10x+42x^2+64x^3+32x^4}{ (s-m_{G}^2)^2 +m_{G}^2 \Gamma_{G}^2}
\nonumber \\
\frac{d\sigma^{(b)}}{dt} &=& \frac{\kappa^4s^2}{1024\pi}
\frac{-x(1+x)(1+2x+2x^2)}{ (s-m_{G}^2)^2 + m_{G}^2\Gamma_{G}^2}\;.
\nonumber \\
\label{angle}
\end{eqnarray}
Here we define as in Ref.~\cite{Giudice:1999ck} $x=t/s$. We set in
accordance with Ref.~\cite{Allanach:2000nr}
$\kappa  = \sqrt{2}x_1 k/({\bar M_{\rm Pl}} m_{G})$, where
$ k/{\bar M_{\rm Pl}} = 0.01$ and $x_1= 3.8317$ is the first
zero of the Bessel function $J_1(x)$ of order 1. Higher zeros of the
Bessel function $J_1(x)$ generate the series of heavy graviton 
resonances that the Randall-Sundrum model predicts. In the analysis
here we will, however, restrict ourself to the first one.
The total width $\Gamma_{G}$ of the spin-2 graviton $G$ 
with mass $m_{G}$ is determined by the sum of
the following partial decay widths \cite{Allanach:2000nr,Han:1999sg}:
\begin{eqnarray}
\Gamma(G \to V\bar V) &=& 
\delta \frac{\kappa^2m_{G}^3}{80\pi} (1-4 r_V)^{1/2}
\left( \frac{13}{12} + \frac{14}{3} r_V + 4 r_V^2\right)
\nonumber \\
\Gamma(G \to f\bar f) &=&
N_c \frac{\kappa^2m_{G}^3}{320\pi} (1-4 r_f)^{3/2}
\left( 1 + \frac{8}{3} r_f\right)
\nonumber \\
\Gamma(G \to gg) &=&
 \frac{\kappa^2m_{G}^3}{20\pi}
\nonumber \\
\Gamma(G \to \gamma\gamma) &=&
 \frac{\kappa^2m_{G}^3}{160\pi} \;.
\end{eqnarray}
$V$ is a massive vector boson ($V=Z^0,W^\pm$) with mass 
$m_V$ and $r_V= m_V^2/m_{G}^2$. For identical particles $\delta=1/2$
and for distinguishable particles $\delta=1$. $f$ is a fermion with mass $m_f$
and $N_c$ its number of colors, if there are any, otherwise 
$N_c=1$. Furthermore, we set  $r_f = m_f^2/m_{G}^2$. 
In case of the width $\Gamma(G \to V\bar V)$
we reproduce the result of Ref.~\cite{Allanach:2000nr}, in all other cases
the ones of Ref.~\cite{Han:1999sg}. 
\section{Event Generator Implementation}

In order to study graviton production in a reasonably realistic 
framework, a single excited graviton $G$ has been introduced to 
the \textsc{Pythia} 6.1 event generator \cite{pythia}. The $G$ 
mass and the dimensionless coupling parameter 
$\kappa m_{G} = \sqrt{2}x_1 k/{\bar M_{\rm Pl}}$
can be set freely. Partial decay widths are given as above, and add
up to a total width used for the resonance Breit-Wigner. The production 
processes $gg \to G$ and $q\bar q \to G$ are included, with 
relevant angular distribution for the subsequent decays of $G$ to a 
fermion pair (while other decays currently are isotropic only). The basic 
process is embedded in the standard \textsc{Pythia} framework of 
initial- and final-state QCD parton showers, underlying event activity 
(multiple interactions and beam remnants), fragmentation to hadrons and 
unstable particle decays. For $G$ decays to lepton pairs, the most 
notable effect may be the initial-state radiation of gluons off the 
incoming quarks and gluons, that gives a $p_{\perp}$ recoil to the
produced $G$.

\section{Simulation of the characteristic $p_\perp$ spectrum}
We present as an example a  simulation for the ATLAS experiment at LHC with
$E_{\rm cm} = 14$ TeV. 
In the following we have to investigate how well
the hypothetical graviton mass can be reconstructed from the 
lepton  pairs.
A comprehensive study using the ATLFAST program \cite{ATLFAST} has
been performed in Ref.~\cite{Allanach:2000nr} for $e^+e^-$ pairs. Here we
want to add the experimental width for $\mu^+\mu^-$ pairs. For the
graviton masses we use the same mass window as in  Ref.~\cite{Allanach:2000nr},
i.e 500 GeV $< m_G <$ 2200 GeV. As  mentioned
in Ref.~\cite{Allanach:2000nr} it will not be possible to detect gravitons
with masses larger than about 2200 GeV at the ATLAS experiment in the
scenario discussed here. On the
other hand, already existing bounds limit the minimum graviton mass
to $m_G >$ 500 GeV \cite{Davoudiasl:2000wi}. It should be noted that
the choice for the coupling constant lies on the lower edge, in fact
the allowed region favors rather $k/\bar M_{\rm Pl} = 0.1$ than
0.01 \cite{Davoudiasl:2000wi}
leading to a cross section which would be two order of magnitudes larger
than the one we have assumed here, but then the graviton resonances
would be no longer narrow. 
In this sense our estimates are conservative.
As we are
only interested in the principle effect we use here an approximative 
parameterization for the resolution of the ATLAS detector. For the electrons 
the following formula is used (see Ref.~\cite{:1999fq} pp. 114-115):  
\begin{equation}
\left(\frac{\sigma(E)}{E}\right)^2_{\rm electrons} \approx  
\frac{(0.1)^2 \;{\rm GeV}}{E} + (0.005)^2\;.
\label{param}
\end{equation}
For the the muons, the combined ATLAS detector resolution for $p_\perp$
measurement, using both the muon spectrometer and the inner tracking 
detectors, is about $2\%$ below $p_\perp=100$ GeV, about
$4\%$ at 300 GeV and about 7\% at 1000 GeV 
(see Ref.~\cite{:1999fq} p.~242).
%
%
%
%%%%%%%%%%%%%%%%%%%%%%%%%%% Figure 2 %%%%%%%%%%%%%%%%%%%%%%%%%%%%%%%%%%%%%%
\begin{figure}[tb]
\begin{center}
\includegraphics[width=8cm]{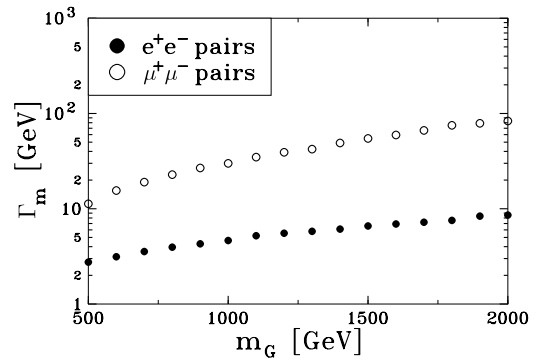}
\end{center}
\caption{Graphical representation of the mass 
resolution $\Gamma_m$ of the ATLAS
detector for narrow graviton resonances with
mass $m_G$ reconstructed from $e^+e^-$ and $\mu^+\mu^-$ pairs
for the ATLAS detector.}
\label{figmassres} 
\end{figure}
%%%%%%%%%%%%%%%%%%%%%%%%%%%%%%%%%%%%%%%%%%%%%%%%%%%%%%%%%%%%%%%%%%%%%%%%%%%
%
%
%
Furthermore, in order to have realistic trigger conditions,
we take $e^+ e^-$ pairs into account only if both
of them have a pseudorapidity $|\eta|<2.5$. Both of them 
must in addition have a transverse energy
larger than 20 GeV, or one of the two electrons has
to have an $E_\perp$ of at least 30 GeV 
(see Ref.~\cite{:1999fq} p.~392).
For the muon pairs we adopt corresponding conditions:
both of them must have a pseudorapidity $|\eta|< 2.4$,  and  both
must have a transverse momentum $p_\perp$ larger than 6 GeV or one of them must
have a $p_\perp$ of at least 20 GeV (see Ref.~\cite{:1999fq} p.~392). 
The number of muon-pair events is thus originally larger
than the number of electron-pair events. The limited detector resolution
leads to a smearing of the reconstructed graviton mass which can be
fitted by a Gaussian distribution. The width of this Gaussian distribution
defines the experimental width $\Gamma_m$. 
Fig.~\ref{figmassres} shows the experimental graviton-mass  
resolution $\Gamma_m$ reconstructed
from $e^+e^-$ and $\mu^+\mu^-$ pairs using the parameterizations given 
above. The values presented here for the $e^+e^-$ pairs agree
roughly with the ones shown in Ref.~\cite{Allanach:2000nr} using the ATLFAST
routine.
%
%
%
%\begin{table}
%\begin{tabular}{|r||c|c|}
%\hline && \\
%$m_G$ [GeV] & $\Gamma_m(e^+e^-)$ [GeV] & $\Gamma_m(\mu^+\mu^-)$ [GeV] \\
% && \\
%\hline
%\hline
% && \\
%500  & 2.752 & 11.27 \\
%600  & 3.134 & 15.53 \\
%700  & 3.555 & 19.03 \\
%800  & 3.950 & 22.81 \\
%900  & 4.284 & 26.79 \\
%1000 & 4.640 & 29.94 \\
%1100 & 5.208 & 34.82 \\
%1200 & 5.534 & 39.13 \\
%1300 & 5.794 & 42.24 \\
%1400 & 6.115 & 49.02 \\
%1500 & 6.595 & 54.68 \\
%1600 & 6.919 & 59.46 \\
%1700 & 7.218 & 66.41 \\
%1800 & 7.590 & 75.11 \\
%1900 & 8.366 & 78.86 \\
%2000 & 8.578 & 83.41 \\
%2100 & 8.680 & 89.02 \\
%2200 & 9.072 & 90.53 \\
% && \\
%\hline
%\end{tabular}
%\caption{Mass resolution $\Gamma_m$ of the ATLAS
%detector for narrow graviton resonances with
%mass $m_G$ reconstructed from $e^+e^-$ and $\mu^+\mu^-$ pairs.}
%\label{tabmassres}
%\end{table} 
%
%
%
%
%
%%%%%%%%%%%%%%%%%%%%%%%%%%% Figure 3 %%%%%%%%%%%%%%%%%%%%%%%%%%%%%%%%%%%%%%
\begin{figure}[tb]
\begin{center}
\includegraphics[width=8cm]{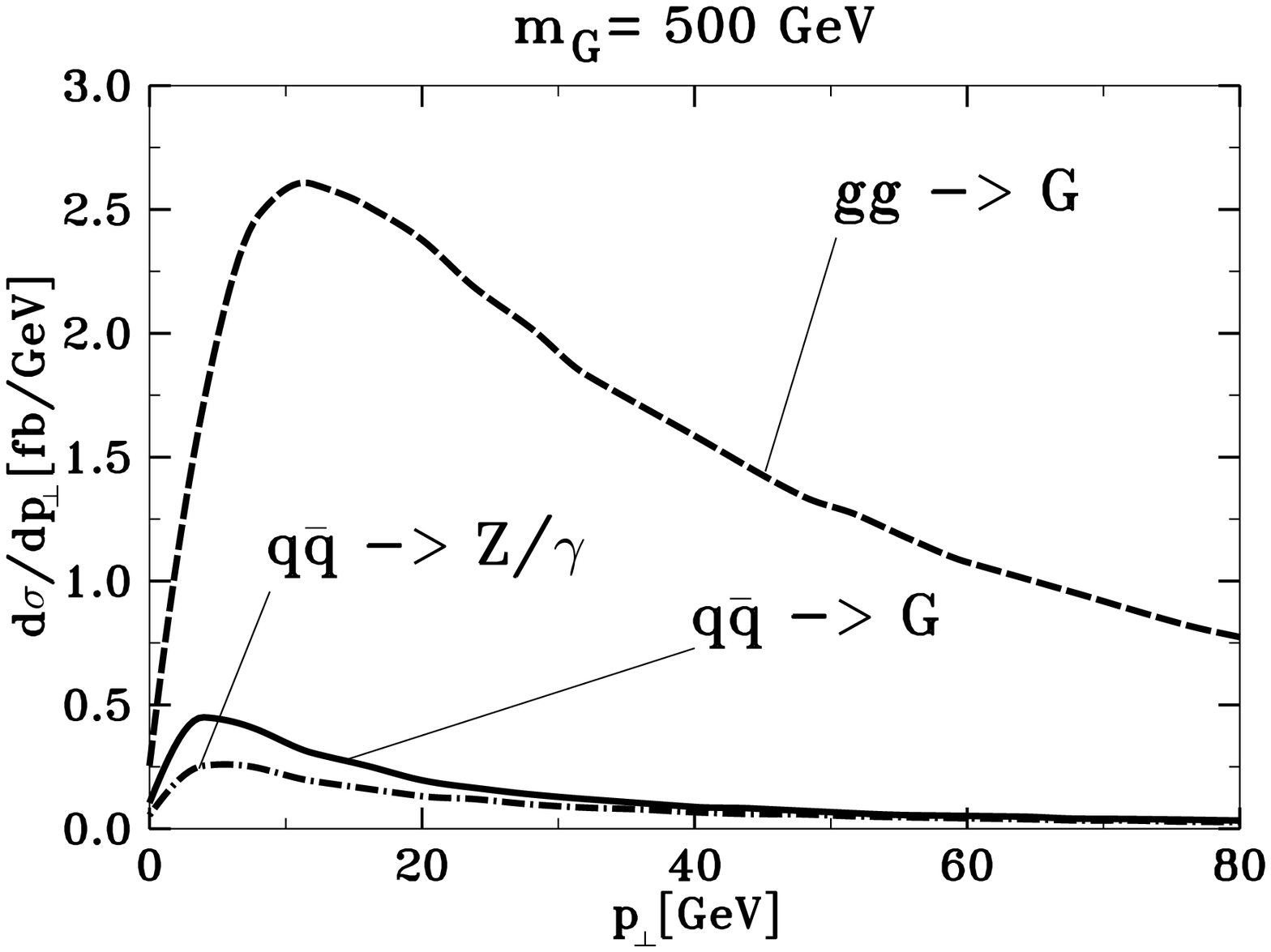}
\includegraphics[width=8cm]{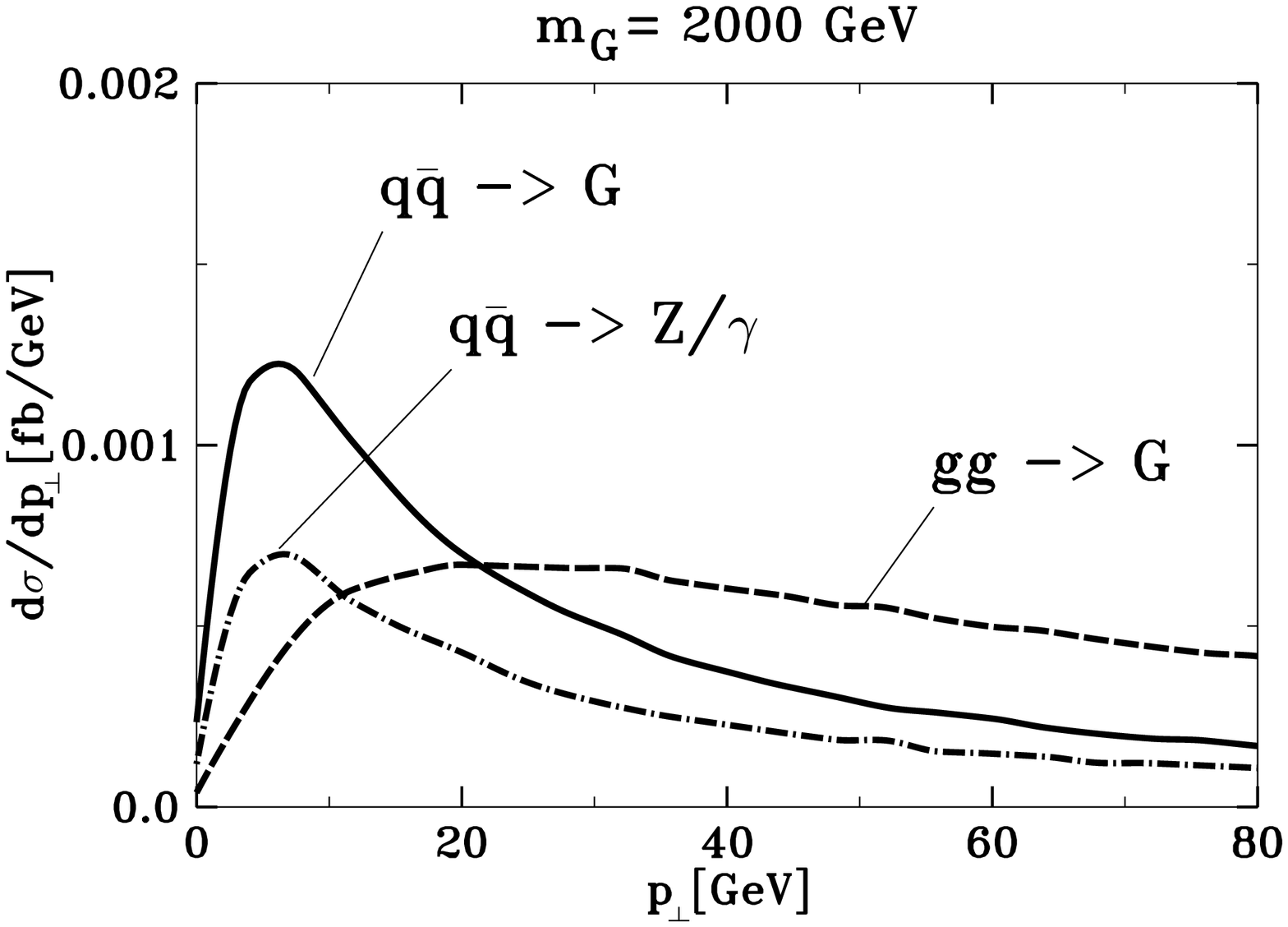}
\end{center}
\caption{$p_\perp$ spectrum for $m_G = 500$ GeV (top) and
$m_G = 2000$ GeV (bottom). In the figure is shown the Standard Model
background from $\gamma,Z^0$ production (dashed-dotted) line, 
the $gg$ contribution (dashed line) and the $q\bar q$ contribution
(solid line).}
\label{ptspec} 
\end{figure}
%%%%%%%%%%%%%%%%%%%%%%%%%%%%%%%%%%%%%%%%%%%%%%%%%%%%%%%%%%%%%%%%%%%%%%%%%%%
%
%
It has to be stated
that in all cases the experimental width exceeds by far the physical width
given by the lifetime of the graviton resonance \cite{Allanach:2000nr}.
But this is inessential
for the forthcoming simulations as the results depend only weakly on the 
precise value of the experimental width and, furthermore, in this paper we only
intend to show the principal effect.
As a result, it is seen that the resolution for the electrons is nearly an
order of magnitude better than for the muons, therefore,
in the following we will concentrate ourselves on the electrons only.
Only in cases where the statistical error will 
overwhelm the experimental resolution the muons might provide additional 
evidence.
As a next step we simulate the $p_\perp$ spectrum for the two hypothetical
graviton masses $m_G = 500$ GeV and $m_G = 2000$ GeV. 
For the simulations we use the width from 
Fig.~\ref{figmassres}. The $p_\perp$ distribution is created from 
initial-state parton showering only. The final-state parton shower complicates
the situation only through photon bremsstrahlung. We have checked that these
latter effects are small. 
\newline
\newline
Fig.~\ref{ptspec} shows the $p_\perp$ spectrum for 
$m_{G} = 500\;{\rm GeV}$ (top) and $m_{G} = 2000\;{\rm  GeV}$ (bottom) split
in the contributions $gg \to G \to e^+e^-$, $q\bar q \to G \to e^+e^-$
and the Standard Model background $q\bar q \to Z/\gamma \to e^+e^-$. To 
reduce the Standard Model background a window of $3\times \Gamma_m$ 
around the 
resonance maximum is taken. It is seen that for smaller resonance masses
the contribution from gg fusion becomes dominant, but in both cases the 
maximum of the $p_\perp$ spectrum for the processes with a $q\bar q$
initial state (including the SM background) lies at considerable
 smaller values than
the one for the gg-fusion process.
Therefore, the characteristic shape of the $p_\perp$ spectrum allows
to draw conclusions about the ratio of $q \bar q$ versus $gg$-processes 
and provides a cross check whether the underlying theoretical model is 
correct. 
%
%
%%%%%%%%%%%%%%%%%%%%%%%%%%% Figure 4 %%%%%%%%%%%%%%%%%%%%%%%%%%%%%%%%%%%%%%
\begin{figure}[tb]
\begin{center}
\includegraphics[width=8cm]{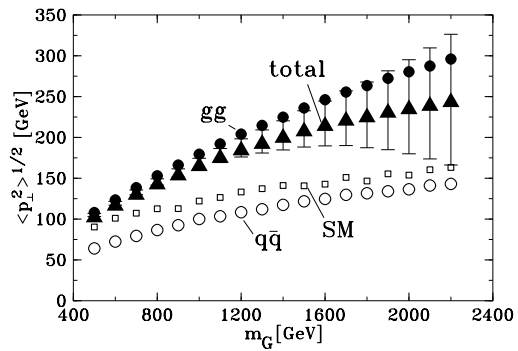}
\end{center}
\caption{Average $p_\perp$, i.e. $\sqrt{\langle p_\perp^2 \rangle}$, 
versus the graviton mass $m_{G}$. The open circles show the average
$p_\perp$ for the $q\bar q \to G$ process, the full circles for 
the the $gg \to G$ process, and the
open squares for the Standard Model background
$q\bar q \to Z/\gamma$. The triangles show the total 
average $p_\perp$ for all processes including the Standard Model
background, with error bars for a luminosity of ${\cal L} = 100\;
{\rm fb}^{-1}$.}
\label{fig3} 
\end{figure}
%%%%%%%%%%%%%%%%%%%%%%%%%%%%%%%%%%%%%%%%%%%%%%%%%%%%%%%%%%%%%%%%%%%%%%%%%%%
%
%
To quantify this we regard in Fig.~\ref{fig3} the average
$p_\perp$, i.e. $\sqrt{\langle p_\perp^2 \rangle}$. The triangles
show the total average $p_\perp$, where all processes including the
SM background contribute. For the errors we use the parameterization
in Eq.~\ref{param} for the electrons plus a statistical
error given by the root of the number of events.
%
%
%\begin{equation}
%\left(\frac{\delta p_\perp}{p_\perp}\right)^2 =
%\frac{1}{{\cal L} \sigma} + 
%\left(5\times 10^{-4} \frac{ p_\perp}{{\rm GeV}}\right)^2
%+ (0.02)^2 \quad.
%\end{equation}
%
%
For the simulation we assume  a luminosity
of ${\cal L}= 100\;{\rm fb}^{-1}$. This is the high luminosity
planned for the ATLAS detector \cite{:1999fq}.
 One sees that even for the highest mass values for the graviton
mass, i.e. $m_{G} = 2200\;{\rm  GeV}$ the pure $q\bar q$ fusion hypothesis
lies outside the 1 $\sigma$ range of the combined $q\bar q$ and $gg$
fusion hypothesis. Furthermore, one observes that for values smaller
than   $m_{G} = 1200\;{\rm  GeV}$ the gg fusion is so dominant
that the total average $p_\perp$ nearly coincides with the one of
gg fusion only. The average $p_\perp$ of the Standard Model background
alone lies a bit above the $q\bar q\to G$ processes. The difference between 
the two $q\bar q$-induced processes mainly comes from the 
$q\bar q \to Z/\gamma$ parton shower being matched to first order matrix 
elements at large $p_\perp$ \cite{pythia}. A careful study of the 
$p_\perp$ spectrum as a whole will be very helpful to distinguish possible 
Randall-Sundrum graviton resonances from pure $q\bar q$ based exotic resonances
like a $Z'$.
%
%
%The average $p_\perp$ as used here can be regarded as the
%square root of second moment of the $p_\perp$ distribution. The
%analysis of even higher moments would unravel more and more characteristic
%details of the spectrum and would allow a supportive statement whether
%a possibly detected resonance follows the originally predicted ratio
%of $q\bar q$ to $gg$ events in the Randall-Sundrum model or not. 
%This would explicitly help to distinguish other exotica such as 
%$Z^*$ bosons which can only be created by $q\bar q$ fusion
%from possible Randall-Sundrum graviton candidates. 
\newline
\newline
The characteristic features of the $p_\perp$ spectrum may
also be at help to  enhance the angular distribution of the
$gg\to G\to e^+e^-$ channel relative to the $q \bar q\to G\to e^+e^-$
channel by cutting out $e^+e^-$ pairs with low $p_\perp$.
The full information content is available in the doubly 
differential distribution of $p_\perp$ and $e^+e^-$ decay angle together 
(Eq.~\ref{angle}). A complete analysis of this issue, e.g. using
a likelihood analysis is, however, beyond the scope of this article.

\section{Matrix elements versus showering formalism}
The $p_\perp$ spectrum described in the previous chapter has been generated
by the parton showering formalism. As this is only an approximation,
it is important to check how well it models the description by a full
matrix element calculation. For this purpose we consider the next
to leading order graviton production matrix elements.
%
%
%
%
%%%%%%%%%%%%%%%%%%%%%%%%%%% Figure 5 %%%%%%%%%%%%%%%%%%%%%%%%%%%%%%%%%%%%%%
\begin{figure}[tb]
\begin{center}
\includegraphics[width=8cm]{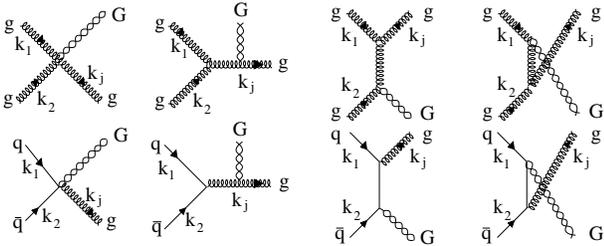}
\end{center}
\caption{NLO amplitudes for the resonance production of gravitons $G$, with
the subprocesses $gg \to Gg$ (top) and $q\bar q \to Gg$ (bottom).}
\label{nlo} 
\end{figure}
%%%%%%%%%%%%%%%%%%%%%%%%%%%%%%%%%%%%%%%%%%%%%%%%%%%%%%%%%%%%%%%%%%%%%%%%%%%
%
%
Fig.~\ref{nlo} shows the NLO contributions to graviton production for
the processes $g(k_1)+ g(k_2) \to g(k_j)+G$ and
$q(k_1)+\bar q(k_2) \to g(k_j)+G$.
For the Mandelstam variables we define $s=(k_1+k_2)^2$, $t=(k_1-k_j)^2$,
and $u=(k_2-k_j)^2$, see Fig.~\ref{mandl}, where $k_j$ denotes
the momentum of the outgoing parton jet. 
Then we have the relation $m_G^2 = s+t+u$. 
In the following we use the shorthand notation $k=k_1+k_2$.
The gluon polarization tensor is noted by $\epsilon^{A\rho}(k_j)$
and the graviton polarization tensor by $\epsilon^{\mu\nu}(k_{gr})$, 
where $k_{gr} = k-k_j$ is the graviton's four-momentum. 
For the processes $q\bar q \to Gg$ the amplitude reads:
\begin{eqnarray}
&& M_{q\bar q \to Gg} = \frac{i}{2} g\kappa \bar v(k_2)t^A\Bigg[
 \frac{1}{t}  \gamma_\mu (\ksla_1-\ksla_j)\gamma_\rho k_{2\nu}
\nonumber \\ &&  \qquad
+\frac{1}{u}  \gamma_\rho (\ksla_2-\ksla_j)\gamma_\mu k_{1\nu}
- \frac{2}{s}\gamma^\sigma  \Bigg( k\cdot k_j g_{\mu\rho}
g_{\nu \sigma} 
\nonumber \\&& \qquad
+ g_{\rho\sigma} k_{j\mu}    k_\nu
- g_{\mu\sigma}  k_{j\nu}    k_\rho
- g_{\mu\rho}    k_{j\sigma} k_\nu \Bigg)
\nonumber \\ && \quad + g_{\mu\rho}\gamma_\nu \Bigg] u(k_1)
\epsilon^{A\rho}(k_j)
\epsilon^{\mu\nu}(k_{gr})\;.
\end{eqnarray}
Here we reproduce 
%up to an convention dependent overall sign 
the results of Ref.~\cite{Han:1999sg}.
For the differential cross section we obtain in agreement with 
Ref.~\cite{Giudice:1999ck}:
\begin{eqnarray}
\frac{d\sigma_{q\bar q \to Gg}}{dt}
&=& \frac{\alpha_s \kappa^2}{64 s} \frac{N_c^2-1}{N_c^2}
\Bigg[
4\frac{t^2+u^2}{s^2} 
+ 9\frac{t+u}{s} 
\nonumber \\ &&
   + \frac{1}{s}\left(\frac{t^2}{u}+\frac{u^2}{t}\right)
   + 3 \left( 4 + \frac{t}{u} + \frac{u}{t}\right)
\nonumber \\ &&
   + 4s \left( \frac{1}{u} + \frac{1}{t}\right)
   + \frac{2s^2}{tu} \Bigg]
\nonumber \\
&=&  \frac{\alpha_s \kappa^2}{64 s^2} \frac{N_c^2-1}{N_c^2} F(s,t,u)
%\nonumber \\
%c_0 &=& 8 \frac{t^2+u^2}{s} \nonumber \\
%c_1 &=& 18(t+u) + 2 \left(\frac{t^2}{u}+\frac{u^2}{t}\right) \nonumber \\
%c_2 &=& 6s \left( 4 + \frac{t}{u} + \frac{u}{t}\right) \nonumber \\
%c_3 &=& 8s^2 \left( \frac{1}{u} + \frac{1}{t}\right) \nonumber \\
%c_4 &=& \frac{4s^3}{tu}\;.
\end{eqnarray}
The cross sections for the processes $qg\to qG$ and $\bar qg \to \bar qG$
can be obtained from this by a simple rotation of the Mandelstam variables:
\begin{equation}
\frac{d\sigma_{qg \to Gq}}{dt} =   
\frac{d\sigma_{\bar q g \to G \bar q}}{dt} =  
\frac{\alpha_s \kappa^2}{64 s^2 N_c}F(t,s,u)\;.
\end{equation}
%
%
%%%%%%%%%%%%%%%%%%%%%%%%%%% Figure 6 %%%%%%%%%%%%%%%%%%%%%%%%%%%%%%%%%%%%%%
\begin{figure}[tb]
\begin{center}
\includegraphics[width=8cm]{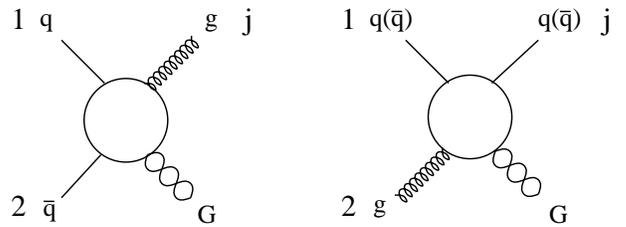}
\end{center}
\caption{Definition of the Mandelstam variables for the processes
$q \bar q \to g G$ (left)  and $q g\to q G$, $\bar q g \to \bar q G$ (right).}
\label{mandl} 
\end{figure}
%%%%%%%%%%%%%%%%%%%%%%%%%%%%%%%%%%%%%%%%%%%%%%%%%%%%%%%%%%%%%%%%%%%%%%%%%%%
%
For the amplitude of the process $gg\to Gg$ one gets using
gauge invariance (where the gauge dependent terms of the inner
gluon propagator vanish identically):
\begin{eqnarray}
&&M_{gg\to Gg}= \frac{\kappa g}{2} f^{ABC}\Bigg\{
                  \frac{1}{t}G_{\alpha\beta}^{\quad\sigma}(k_j,-k_1,k_1-k_j)
\nonumber \\ && \quad \quad \times
[k_2\cdot(k_1-k_j) C_{\mu\nu,\rho\sigma}+ D_{\mu\nu,\rho\sigma}(k_2,k_1-k_j)]
\nonumber \\ && \;\;
+ \frac{1}{u}G_{\rho\alpha}^{\quad\sigma}(-k_2,k_j,k_2-k_j)
\nonumber \\ && \quad \quad\times
[k_1\cdot(k_2-k_j) C_{\mu\nu,\beta\sigma}+ D_{\mu\nu,\beta\sigma}(k_1,k_2-k_j)]
\nonumber \\ && \;\;
+ \frac{1}{s}G_{\beta\rho}^{\quad\sigma}(-k_1, -k_2,k_1+k_2)
\nonumber \\ && \quad \quad \times
[-k_j\cdot(k_1+k_2) C_{\mu\nu,\alpha\sigma}+ D_{\mu\nu,\alpha\sigma}(-k_j,k_1+k_2)]
\nonumber \\ && \;\; 
+[ C_{\mu\nu,\rho\beta}(k_2-k_1)_\alpha
+ C_{\mu\nu,\rho\alpha}(-k_j-k_2)_\beta
\nonumber \\ && \;\;\;
+ C_{\mu\nu,\beta \alpha}(k_1+k_j)_\rho
+ F_{\mu\nu,\rho\beta \alpha}(k_2,k_1,-k_j)]\Bigg\}
\nonumber\\&&  \times \epsilon^{\alpha A}(k_j)\epsilon^{\beta B}(k_1)
\epsilon^{\rho C}(k_2) \epsilon^{\mu\nu}(k_{gr})\;.
\end{eqnarray}
Here the following definitions are used:
\begin{eqnarray}
C_{\mu\nu,\rho\sigma} &=& 2 g_{\mu\sigma}g_{\nu \rho}
\nonumber  \\
D_{\mu\nu,\rho\sigma}(k_1,k_2) &=& 
-2(
     g_{\mu\sigma } k_{1\nu}   k_{2\rho}
  +  g_{\mu\rho   } k_{1\sigma}k_{2\nu}
\nonumber \\ && \quad
  -  g_{\rho\sigma} k_{1\mu}   k_{2\nu})
\nonumber \\
F_{\mu\nu,\rho\sigma\lambda}(k_1,k_2,k_3) &=&
 2(g_{\mu \lambda} g_{\rho   \sigma}(k_1-k_2)_\nu
\nonumber \\ && \quad
  +g_{\mu \rho   } g_{\sigma \lambda}(k_2-k_3)_\nu
\nonumber \\ && \quad
  +g_{\mu \sigma } g_{\lambda \rho}(k_3-k_1)_\nu)
\nonumber \\
G_{\rho\sigma\lambda}(k_1,k_2,k_3) &=& 
  g_{\rho   \sigma}(k_1-k_2)_\lambda
+ g_{\sigma \lambda}(k_2-k_3)_\rho
\nonumber \\ && 
+ g_{\lambda \rho}(k_3-k_1)_\sigma\;.
\end{eqnarray}
Furthermore, for the spin-sums over the polarization tensors one gets:
\begin{eqnarray}
\sum_{s=1}^5 \epsilon_{\mu\nu}^{s}(k_g)\epsilon_{\rho\sigma}^{*s}(k_g)
&=&            \tilde \eta_{\mu\rho}   \tilde \eta_{\nu\sigma} 
             + \tilde \eta_{\mu\sigma} \tilde \eta_{\nu\rho} 
  -\frac{2}{3} \tilde \eta_{\mu\nu}    \tilde \eta_{\rho\sigma}
\nonumber \\
\tilde \eta_{\mu\nu} &=& g_{\mu\nu} - \frac{k_{g\mu}k_{g\mu}}{m_G^2}
\nonumber \\
\sum_{s=1}^3 \epsilon_{\mu}^{sA}(k_i)\epsilon_{\nu}^{*sB}(k_i) 
&=& \delta^{AB} \left( -g_{\mu\nu} 
+ \frac{k_{i\mu} n_{i\nu}+ k_{i\nu} n_{i\mu}}{k_i\cdot n_i}\right) 
\nonumber \\
n_i\cdot n_i &=& 0;\quad (i=1,2,j)\;.
\end{eqnarray} 
For the partonic cross section for the process $gg\to Gg$ we
obtain then in agreement with Ref.~\cite{Giudice:1999ck}:
\begin{eqnarray}
\frac{d\sigma_{gg\to Gg}}{dt} &=& \frac{N_c\kappa^2\alpha_s}{4s^2(N_c^2-1)}
\Bigg[ - 4(s+t+u)
\nonumber \\
&& + 
\frac{(s^2+t^2+u^2 +st+su+tu)^2}{stu} \Bigg]\;,
\end{eqnarray}
%
%
%\begin{eqnarray}
%\frac{d\sigma_{gg\to Gg}}{dt} &=& \frac{3\kappa^2\alpha_s}{s(N_c^2-1)^2}
%\Bigg[ 2 \frac{(t^2+tu+u^2)^2}{s^2tu}
%      + \frac{4}{s}\left(\frac{t^2}{u}+\frac{u^2}{t}\right)
%\nonumber \\
%&& +  6\left( \frac{t}{u} + \frac{u}{t}\right)
%   + 4s \left( \frac{1}{u} + \frac{1}{t}\right)
%   +  \frac{2s^2}{tu} \Bigg]
% \nonumber \\
%cm1 &=& 16 \frac{(t^2+tu+u^2)^2}{stu} \nonumber \\
%c_0 &=&  32 \left(\frac{t^2}{u}+\frac{u^2}{t}\right) \nonumber \\
%c_1 &=& 48s\left( \frac{t}{u} + \frac{u}{t}\right) \nonumber \\
%c_2 &=& 32s^2 \left( \frac{1}{u} + \frac{1}{t}\right) \nonumber \\
%c_3 &=& \frac{16s^3}{tu} \nonumber \\
%dPS &=&  \frac{dt}{16\pi s^2}\;.
%\end{eqnarray}
%
%
which shows the symmetry under the exchange of all three gluons
with each other.
These cross sections are implemented into the event generator
\textsc{Pythia} 6.1 as well. 
Next we consider the ratio of the NLO matrix elements
versus the LO matrix elements plus parton showering. For $p_\perp$ values
larger than $100 \;{\rm GeV}$ where soft effects and NNLO contributions
are negligible this ratio should actually become equal to one in a certain
range. Below $100 \;{\rm GeV}$ a resummation procedure would have
to be applied to tame the $p_\perp\to 0$ divergence of the NLO matrix
elements, as is already implicit in the shower formalism.
The situation is slightly complicated by the fact that the basic
subprocesses that initiate the parton showering processes are only
$q\bar q \to G$ and $gg \to G$, see Fig.~\ref{graphs}, whereas the NLO
matrix elements contain processes with $qg$ and $\bar q g$ initial parton
states as well. 
\newline
\newline
%%%%%%%%%%%%%%%%%%%%%%%%%%% Figure 7 %%%%%%%%%%%%%%%%%%%%%%%%%%%%%%%%%%%%%%
\begin{figure}[tb]
\begin{center}
\includegraphics[width=8cm]{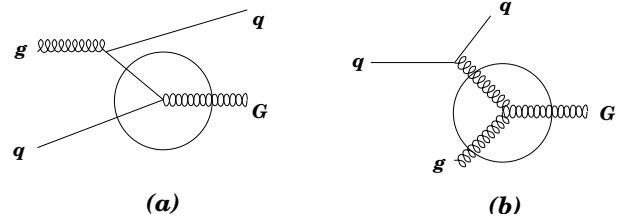}
\end{center}
\caption{Assignment of $qg\to qG$ ($\bar qg\to \bar qG$)
 processes in the parton shower formalism.
The figure shows a  $qg\to qG$ process containing a $q\bar q \to G$ vertex
(a) and a $gg \to G$ vertex (b).}
\label{qgproblem} 
\end{figure}
%%%%%%%%%%%%%%%%%%%%%%%%%%%%%%%%%%%%%%%%%%%%%%%%%%%%%%%%%%%%%%%%%%%%%%%%%%%
%
The shower branchings effectively induce such
initial states, see Fig.~\ref{qgproblem}, so there is no fundamental conflict, 
but more a practical issue of comparing different classification schemes.
In general, one would have to share the $qg/ \bar qg$ NLO  contributions 
between the $q\bar q$ and the $gg$ shower processes. 
%
%
%%%%%%%%%%%%%%%%%%%%%%%%%%% Figure 8 %%%%%%%%%%%%%%%%%%%%%%%%%%%%%%%%%%%%%%
\begin{figure}[tb]
\begin{center}
\includegraphics[width=8cm]{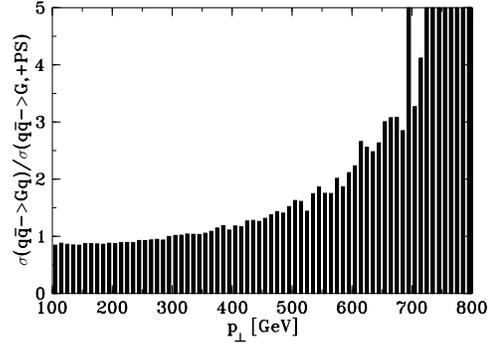}
\end{center}
\caption{Comparison of matrix elements versus parton showering 
for the $q\bar q$ graviton production.
Displayed is ratio of the NLO $q\bar q \to Gg$ cross section versus the 
LO cross section $q\bar q \to G$ plus parton showering.}
\label{nloqq} 
\end{figure}
%%%%%%%%%%%%%%%%%%%%%%%%%%%%%%%%%%%%%%%%%%%%%%%%%%%%%%%%%%%%%%%%%%%%%%%%%%%
%
%
We note however,
that a $qg\to qG$ graph with a $gg \to G$ vertex (Fig.~\ref{qgproblem}b)
would receive contributions from t-channel gluon exchange, 
while the same graph with a $q\bar q \to G$ vertex (Fig.~\ref{qgproblem}a) 
would instead 
contain u-channel quark exchange. 
The fact that the $qg\to qG$ cross section
is strongly peaked at small $t$, and not at small $u$, indicates
that $qg\to qG$ predominantly contributes to the $gg\to G$ graph and only 
little to  $q\bar q \to G$.
Fig.~\ref{nloqq} shows the ratio
$\sigma(q\bar q\to Gg)/\sigma(q\bar q\to G + PS)$ versus $p_\perp$. It
is seen that for $p_\perp$ lower than $400$ GeV we find a good
 agreement between NLO matrix elements and 
the LO+parton shower, as the ratio here is nearly equal to one, as predicted
by the reasoning above.
\newline\newline
%
%
%
%
%%%%%%%%%%%%%%%%%%%%%%%%%%% Figure 9 %%%%%%%%%%%%%%%%%%%%%%%%%%%%%%%%%%%%%%
\begin{figure}[tb]
\begin{center}
\includegraphics[width=8cm]{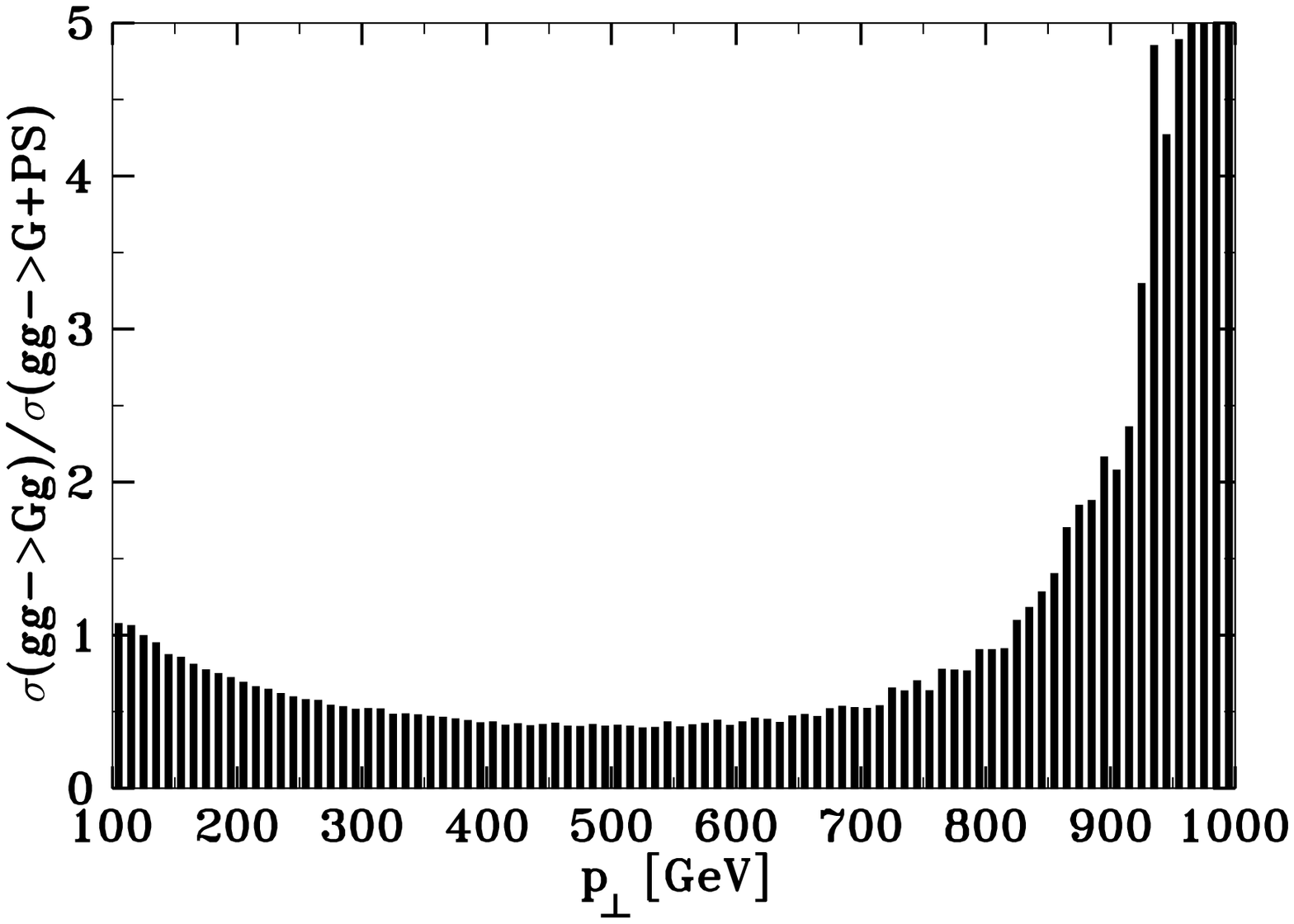}
\includegraphics[width=8cm]{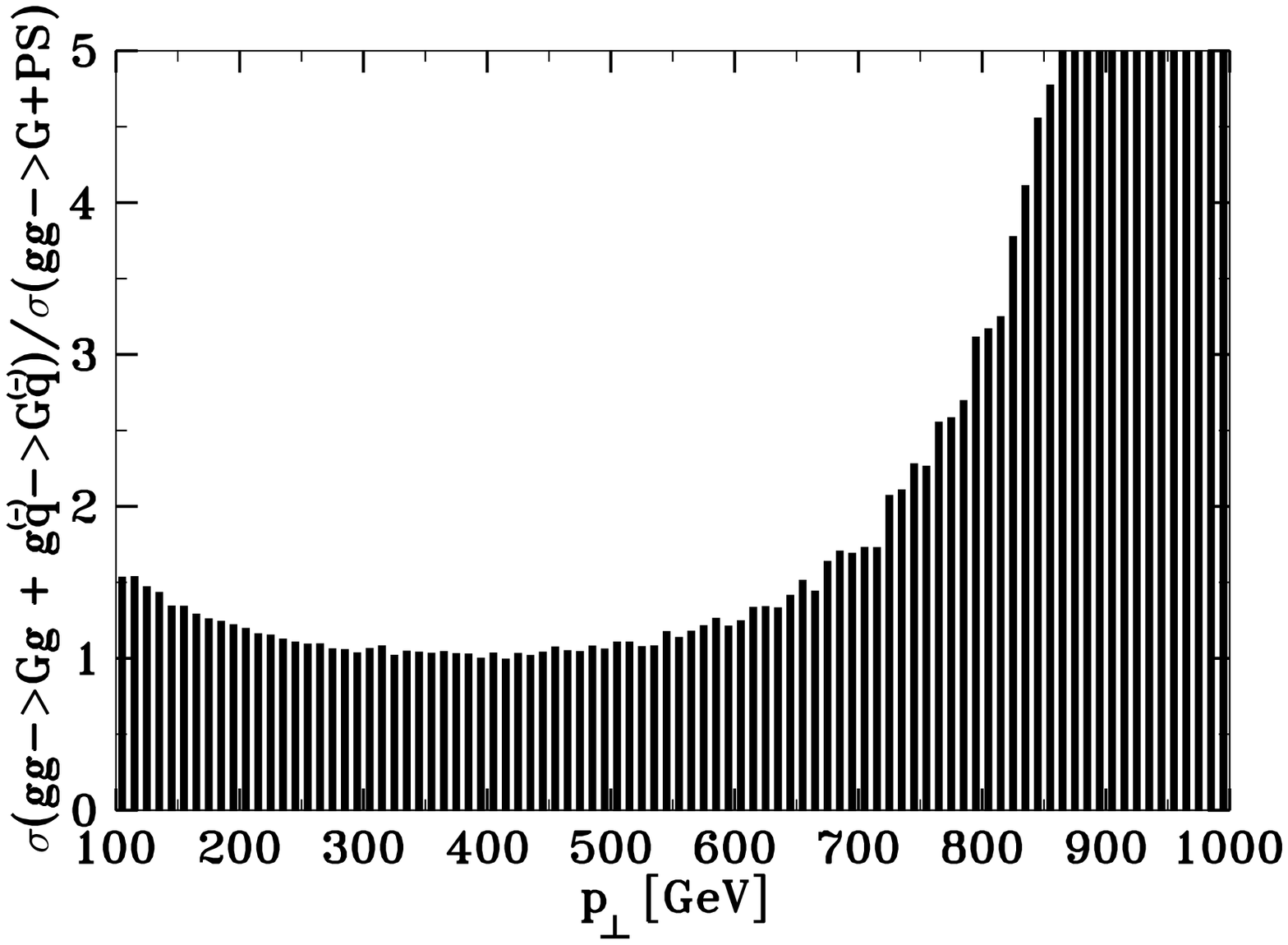}
\end{center}
\caption{Comparison of matrix elements versus parton showering 
for graviton production involving gluons. Displayed is the 
ratio $\sigma(gg\to Gg)/\sigma(gg\to G + PS)$ (top) and
$(\sigma(gg\to Gg)+\sigma(gq\to Gq) + \sigma(g\bar q \to G\bar q))
/\sigma(gg\to G + PS)$ (bottom).}
\label{nlogg} 
\end{figure}
%%%%%%%%%%%%%%%%%%%%%%%%%%%%%%%%%%%%%%%%%%%%%%%%%%%%%%%%%%%%%%%%%%%%%%%%%%%
%
Fig.~\ref{nlogg} shows that it is also 
reasonable to assign the $qg/\bar qg$ mixed initial states to
the $gg\to G$ plus shower processes: If
we only consider the ratio $\sigma(gg\to Gg)/\sigma(gg G + PS)$
one sees that in the range 100 GeV $<p_\perp<$ 400 GeV the matrix
elements account for only 50\% of the whole cross section 
coming from parton showering at some values of $p_\perp$. 
If one adds the quark-gluon matrix elements,
however, the ratio is nearly equal to one up to 600 GeV. Therefore we find
that the parton showering formalism is in good agreement with the
cross section given by NLO matrix elements in the $p_\perp$ range
between 100 GeV and 400 GeV. Below 100 GeV the shower formalism should 
give a trustworthy $p_\perp$ spectrum. Then the range we need for the 
analysis of the narrow graviton resonances is covered,  see Fig.~\ref{fig3}.
\indent

\section{Summary}
The $p_\perp$ spectrum is a supportive signature for the 
detection of narrow graviton resonances at LHC. It gives
additional hints on the underlying production processes and
may help to verify or to exclude certain scenarios such as 
the Randall-Sundrum model, because it is sensitive to  a characteristic
mixture of $gg$ and $q\bar q$ processes in graviton production
unique for the corresponding model. Furthermore, we have shown that the parton
showering formalism at TeV collider energies  still gives a correct
approximation of the predictions of matrix element calculations, so that
the approximations in  the parton showering formalism are justified also 
in this kind of processes yet experimentally untested.


\vspace{-0.1cm}
\begin{thebibliography}{99}

%\cite{Randall:1999ee}:
\bibitem{Randall:1999ee}
L.~Randall and R.~Sundrum,
%``A large mass hierarchy from a small extra dimension,''
Phys.\ Rev.\ Lett.\  {\bf 83}, 3370 (1999)
[hep-ph/9905221].
%%CITATION = HEP-PH 9905221;%%



%\cite{Arkani-Hamed:1998rs}
\bibitem{Arkani-Hamed:1998rs}
N.~Arkani-Hamed, S.~Dimopoulos and G.~Dvali,
%``The hierarchy problem and new dimensions at a millimeter,''
Phys.\ Lett.\  {\bf B429}, 263 (1998)
[hep-ph/9803315].
%%CITATION = HEP-PH 9803315;%%

%\cite{Antoniadis:2000vd}:
\bibitem{Antoniadis:2000vd}
I.~Antoniadis and K.~Benakli,
%``Large dimensions and string physics in future colliders,''
hep-ph/0007226.
%%CITATION = HEP-PH 0007226;%%

%\cite{Hoyle:2000cv}:
\bibitem{Hoyle:2000cv}
C.~D.~Hoyle, U.~Schmidt, B.~R.~Heckel, E.~G.~Adelberger, J.~H.~Gundlach, D.~J.~Kapner and H.~E.~Swanson,
%``Sub-millimeter tests of the gravitational inverse-square law: A search for "large" extra dimensions,''
hep-ph/0011014.
%%CITATION = HEP-PH 0011014;%%


%\cite{Allanach:2000nr}:
\bibitem{Allanach:2000nr}
B.~C.~Allanach, K.~Odagiri, M.~A.~Parker and B.~R.~Webber,
%``Searching for narrow graviton resonances with the ATLAS detector at the  Large Hadron Collider,''
JHEP {\bf 0009}, 019 (2000)
[hep-ph/0006114].
%%CITATION = HEP-PH 0006114;%%

%\cite{Giudice:1999ck}:
\bibitem{Giudice:1999ck}
G.~F.~Giudice, R.~Rattazzi and J.~D.~Wells,
%``Quantum gravity and extra dimensions at high-energy colliders,''
Nucl.\ Phys.\  {\bf B544}, 3 (1999)
[hep-ph/9811291].
%%CITATION = HEP-PH 9811291;%%

%\cite{Han:1999sg}:
\bibitem{Han:1999sg}
T.~Han, J.~D.~Lykken and R.~Zhang,
%``On Kaluza-Klein states from large extra dimensions,''
Phys.\ Rev.\  {\bf D59}, 105006 (1999)
[hep-ph/9811350].
%%CITATION = HEP-PH 9811350;%%


\bibitem{pythia}
T. Sj\"ostrand, P. Ed\'en, C. Friberg, L. L\"onnblad, G. Miu, 
S. Mrenna and E. Norrbin , LU TP 00--30 [hep-ph/0010017], to
appear in Computer Phys. Commun.


\bibitem{ATLFAST} E.~Richter-Was, D.~Froidevaux and L.~Poggioli, ``ATLFAST
1.0 A package for particle-level analysis'', ATLAS Internal Notes
ATL-PHYS-96-079 (1996) and ATL-PHY-98-131 (1998)

%\cite{Davoudiasl:2000wi}
\bibitem{Davoudiasl:2000wi}
H.~Davoudiasl, J.~L.~Hewett and T.~G.~Rizzo,
%``Experimental probes of localized gravity: On and off the wall,''
hep-ph/0006041.
%%CITATION = HEP-PH 0006041;%%


%\cite{:1999fq}:
\bibitem{:1999fq}
``ATLAS Collaboration: ATLAS detector and physics performance technical 
design report. Volume 1,''
CERN-LHCC-99-14.




\end{thebibliography}
\end{document}